\documentclass[journal]{IEEEtran}
\usepackage{graphicx}
\usepackage{multirow}
\usepackage{amsfonts}
\usepackage{amsmath}
\usepackage{stfloats}
\usepackage{color}
\usepackage{graphicx} 
\usepackage{algorithm}
\usepackage{algorithmic}
\usepackage{verbatim}
\usepackage{float}
\usepackage{cite}
\usepackage{hyperref}

\begin{document}
\title{Uplink SCMA-empowered Uncoordinated Random Access for Future mMTC}

\author{Pengyu Gao, Qu Luo \textit{Member, IEEE}, Jing Zhu \textit{Member, IEEE}, Gaojie Chen \textit{Senior Member, IEEE}, Pei Xiao \textit{Senior Member, IEEE}, and  Chuan Heng Foh, \textit{Senior Member, IEEE}

\thanks{This work of Jing Zhu was supported by the China Postdoctoral Science Foundation under Grant number 2025M773513. This work of Pei Xiao and Qu Luo was supported  by the U.K. Engineering and Physical Sciences  Research  Council  under Grant EP/P03456X/1 and  EP/X013162/1.(corresponding author: Qu Luo) }

\thanks{Pengyu Gao is affiliated with the Department of Broadband Communications, Peng Cheng Laboratory, Shenzhen, Guangdong 518055, China (e-mail: gaopy@pcl.ac.cn).}

\thanks{Qu Luo, Pei Xiao and Chuan Heng Foh are affiliated with 5G and 6G Innovation
Centre, Institute for Communication Systems (ICS) of University of Surrey,
Guildford, GU2 7XH, UK (e-mail: {q.u.luo, p.xiao, c.foh}@surrey.ac.uk)}

\thanks{
Jing Zhu and Gaojie Chen are affiliated with the School of Flexible
Electronics (SoFE) \& State Key Laboratory of Optoelectronic Materials
and Technologies (OEMT), Sun Yat-sen University, Shenzhen, Guangdong
518107, China (e-mail: ZJ009944@hotmail.com, gaojie.chen@ieee.org).}
}%

\maketitle

\begin{abstract}
In this paper, a novel uncoordinated random access (URA) protocol is presented to address the pressing demand for massive connectivity with low access latency in future massive machine type communication (mMTC) scenarios. The proposed URA scheme integrates the classical slotted ALOHA (S-ALOHA) protocol with sparse code multiple access (SCMA) technique, referred to as SCMA-empowered URA. Specifically, active users randomly choose an SCMA codebook to access the communication network in an arbitrary time slot whenever they want without scheduling. However, due to the lack of central coordination in the proposed URA scheme, SCMA codebook collisions become inevitable, making decoding challenging and leading to increased access failures. To cope with the decoding issue, an interference-canceling (IC) first decoding strategy is proposed at the access point (AP), which can partially tackles collision problems, contributing to a higher system throughput. Taking the proposed IC-first decoding strategy into account, a closed-form theoretical expression of the throughput is derived. Moreover, to alleviate the throughput degradation under the congested user traffic, a user barring mechanism is introduced to manage the traffic load. Firstly, a closed-form expression of idle codebook probability is developed to help indicate the system state, i.e., congested or not. Then, in addition to the estimated real-time load, the AP adaptively adjusts the access probability and redistributes the actual access load. Finally, simulation results demonstrate that the proposed SCMA-empowered URA scheme enjoys higher maximum throughput, compared to the conventional orthogonal multiple access (OMA) based URA scheme. Moreover, the accuracy of the presented theoretical analysis and the effectiveness of the user barring mechanism are verified.

\end{abstract}

\begin{IEEEkeywords}
Massive machine type communication (mMTC), sparse code multiple access (SCMA), uncoordinated random access (URA), interference cancellation (IC), theoretical analysis, user barring design.
\end{IEEEkeywords}

\IEEEpeerreviewmaketitle

\section{Introduction}

\IEEEPARstart{W}{ith} the advent of fifth-generation (5G) and beyond, the rapid growth of Internet of Things (IoT) has ushered in an era of unprecedented connectivity demands \cite{IoT-1}, \cite{IoT-2}. Different from the legacy human-centric contexts, massive machine-type communication (mMTC) in future IoT entails the interconnection of an enormous number of devices and machines, encompassing a wide range of applications, from smart cities and industrial automation to health-care and agriculture \cite{mMTC-1}. The traffic characteristics in mMTC are summarized as uplink-dominated, sporadic transmission, small data size, low-power consumption and so on \cite{mMTC-4}. A challenging issue inside is how to meet the requirements of massive connectivity and strict latency constraints, which has aroused extensive research attentions from both academia and industry \cite{mMTC-5}.

To support massive connectivity, promising multiple access techniques are indispensable. From second generation (2G) systems to fourth generation (4G) systems, orthogonal multiple access (OMA) techniques have been widely used, where available orthogonal resources are exclusively allocated to each user \cite{mMTC-5}. However, since OMA can only support limited number of users because of the finite orthogonal resources, it may be infeasible in mMTC scenarios. As an alternative, non-orthogonal multiple access (NOMA) is employed to provide high spectrum efficiency and enable massive connectivity by allowing multiple users simultaneously share the same resources \cite{NOMA-0}. There are two major types of the existing NOMA techniques: power-domain NOMA (PD-NOMA) and code-main NOMA (CD-NOMA) \cite{NOMA-1}. Particularly, PD-NOMA enables multiple users to share the same resource with distinct power levels \cite{NOMA-2}, while CD-NOMA supports multiplexing by assigning individual signature codes to different users \cite{NOMA-3}. The research focus of this paper is on a representative spreading-based CD-NOMA scheme, termed as sparse code multiple access (SCMA) \cite{SCMA1-1}, \cite{SCMA1-2}. In SCMA, each user's bit information is directly mapped to a multi-dimensional sparse codeword, which is selected from a pre-designed codebook. At the receiver side, an efficient message passing algorithm (MPA) is implemented, enabling near-optimum decoding performance owing to the exploitation of the inherent sparsity of SCMA codewords \cite{SCMA1-3}. Moreover, benefiting from the dedicated codebook design, SCMA enjoys a specific shaping gain, making it as an attractive choice in mMTC scenarios \cite{SCMA1-4}.

In addition to the requirement of massive connectivity, low access latency is another critical concern \cite{mMTC-4}. In the long-term-evolution (LTE) standard, the classical four-step handshaking procedure has been adopted \cite{RA-1}-\cite{SCMA1-6}. As a grant-based random access (RA) protocol, it involves two-round data exchange between the access point (AP) and IoT users, namely preamble transmission, random access response, connection request and contention resolution. In the context of mMTC, such complicated handshaking procedure is prohibitively costly, leading to excessive signaling overhead and high access latency \cite{mMTC-5}. Moreover, considering the rapid growth in the number of users, severe preamble collisions are inevitable due to the limited physical resources, thus resulting in the access performance degradation \cite{SCMA1-6}. Against this background, grant-free RA protocols have been introduced, which permit users to access the communication systems without waiting for any requests \cite{mMTC-4}, \cite{gf-1}-\cite{SCMA-1}. Hence, by simplifying the request-grant procedure, access latency can be significantly reduced. Considering the advantages of the NOMA technique, the synergistic combination of SCMA and grant-free RA protocol can be explored as a promising technique in mMTC scenarios to support massive uplink transmission, while maintaining low access latency \cite{SCMA-1}.

Numerous researech works have been conducted on the uplink grant-free SCMA system. In \cite{SCMA-2}, a contention transmission unit (CTU) was defined for the uplink grant-free SCMA transmission, which combines time, frequency, SCMA codebook and pilot. Moreover, \cite{SCMA-3} proposed a joint MPA (JMPA) algorithm to address the blind detection problem, aiming at jointly finding out active users and decoding the corresponding signals. Liu \emph{et al}. in \cite{SCMA-4} introduced compressed sensing (CS) based algorithms to identify active users at the grant-free SCMA receiver, without prior knowledge of the user activity. By further modifying the JMPA method, a low false detection probability was achieved. Additionally, the authors in \cite{SCMA-5} developed a novel SCMA codewords, which is able to distinguish the signal of active/inactive users.  Authors of \cite{SCMA-6} proposed a compressive sensing based SCMA system, where active users are identified by a two-step procedure and the corresponding data symbols are detected by MPA-based detector. In \cite{SCMA-7}, an expectation propagation aided messaging-passing receiver was devised for uplink grant-free SCMA systems, which targets at jointly carrying out channel estimation, user activity detection and data decoding. Then, \cite{SCMA-8} designed an autoencoder-based framework that jointly optimizes preamble generation and data-aided active user detection through a novel user activity extraction network for grant-free SCMA systems. Furthermore, a deep learning-based blind multiuser detector for uplink grant-free SCMA system proposed by \cite{SCMA-9} can adaptively identify the active users with a high reliability, even when the user sparsity level is large. All the above works studied the contention-free grant-free RA mode, where each user is pre-allocated with a unique sequence resource and resource collisions are ignored. However, in this case, huge available physical resources serving for uplink access are required and the number of potential users are assumed to be known in advance and be fixed, which may be unrealistic \cite{NOMA-0}.

Alternatively, the uncoordinated RA (URA) protocol, following the contention-based grant-free RA mode, is considered, where multiple IoT users compete for the limited physical resources for uplink access without the prior coordination and resource pre-allocation. Notably, as a typical URA protocol, slotted ALOHA (S-ALOHA) has garnered significant interest due to easy implementation \cite{SA-1}, \cite{SA-2}, where these users select a single time slot for uplink transmissions whenever they have data to send. In other words, this approach enables an "arrive-and-go" manner of uplink access, with users transmitting their data upon arrival. To the best of our knowledge, studies about the synergistic integration of SCMA are URA are very limited, although the integration of URA and PD-NOMA has been extensively investigated \cite{NOSA-1}-\cite{NOSA-4}. In this paper, we propose an SCMA-empowered URA scheme, where each active IoT user randomly selects an SCMA codebook to map their data information and subsequently delivers to the AP in each transmission time slot. Since SCMA can accommodate more users simultaneously than the number of available physical resources, the maximum system throughput of the proposed scheme achieves evident improvement\footnotemark[1], compared with that of the conventional OMA-based URA scheme. In addition, thanks to the property of URA, the proposed scheme enjoys low access latency as well as low signaling overhead, which is well-suited for mMTC.

Unlike contention-free grant-free RA protocols, the URA protocol faces a critical challenge due to the lack of central coordination and resource pre-allocation. This absence of coordination inevitably leads to resource collisions, which significantly degrade system performance, particularly when a large number of users attempt to access the network simultaneously. To mitigate the adverse impact of resource collisions, user barring mechanisms are introduced to control the real-time user traffic load \cite{Bar-0}, which have been explored by prior studies. Duan \textit{et al.} in \cite{Bar-1} proposed a dynamic access control method to update the access probability for RA delay reduction by taking into account both the number of active users and the number of preamble collisions at the previous time slots. Moreover, considering the service difference, \cite{Bar-2} conceived a quality of service (QoS)-based access control mechanism, which assigns different access probabilities to delay-sensitive and delay-tolerant users. In context of PD-NOMA URA systems, authors of \cite{Bar-3} considered a Q-learning based access algorithm, while \cite{Bar-4} introduced a multi-armed bandit (MAB) enabled method to manage the user access. Besides machine learning (ML) based access control method, \cite{Bar-5} regulated the time-varying traffic load according to a Markov decision process for PD-NONA URA. In this paper, we design an access control mechanism to alleviate codebook collisions, specifically tailored for the proposed SCMA-empowered URA schemes. In particular, the access probability is adaptively determined by estimating the real-time traffic load. The core of our traffic load estimation lies in leveraging the presented theoretical expressions of the system throughput and the idle codebook probability. Thus, by adaptively adjusting the access probability, the optimal traffic load can be achieved, maintaining the system throughput steadily close to the maximum value under the varying traffic load.

\footnotetext[1]{The system throughput in this paper is defined as the number of users whose transmitted packets can be successfully decoded.}

\emph{The main contributions of this paper are summarized as follows:}

\begin{itemize}
\item{In this study, we propose a novel uplink URA scheme integrated with SCMA, denoted as SCMA-empowered URA. Firstly, leveraging the inherent advantages of URA, the proposed system eliminates complex signaling handshakes, achieving low access latency. Then, by incorporating the strengths of SCMA, the system gains the capability to support a much higher number of concurrent users, leading to a significant improvement in system throughput. Furthermore, this increased capacity leads to fewer failed access attempts and consequently fewer retransmissions, further decreasing the overall system access latency.}
\end{itemize}

\begin{itemize}
\item{To resolve the decoding challenges arising from codebook collisions, an interference-cancellation (IC) first decoding strategy is developed, which can partially resolve the collision issue, thereby enhancing the system throughput. Moreover, thanks to the distinctive sparse structure of SCMA codebook, the implementation of the IC process is facilitated, resulting in a low decoding complexity.}
\end{itemize}

\begin{itemize}
\item{We undertake a systematic theoretical analysis of our proposed SCMA-empowered URA scheme. Firstly, we derive a theoretical closed-form expression by fully incorporating our proposed IC-first blind decoding capability, which is tailored for codebook collisions. Next, the idle codebook probability is investigated, which helps the real-time traffic load estimation due to its monotonic decrease property with respect to the traffic load. Its corresponding closed-form expression is presented.}
\end{itemize}

\begin{itemize}
\item{To manage congestion and mitigate the throughput degradation in the proposed scheme, we develop a user barring mechanism, which regulates real-time traffic load by first estimating it using the derived theoretical expressions of system throughput and idle codebook probability. Based on the results of the traffic load estimation, the user barring mechanism adjusts access requests from active IoT users to maintain system throughput close to its maximum value, especially under heavy traffic conditions.}
\end{itemize}

\begin{itemize}
\item{Simulation results demonstrate that 1) our derived closed-form expression of the theoretical system throughput concurs with the exact throughput when utilizing the proposed IC-first decoding strategy; 2) the proposed closed-form expression of the idle codebook probability is accurate; and 3) the proposed user barring scheme is validated to be effective, especially in scenarios where a great number of IoT users simultaneously access the system.}
\end{itemize}

The rest of this paper is organized as follows. The system model is presented in Section II. Next, Section III introduces the proposed IC-first blind decoding strategy. The theoretical analysis of our proposed scheme is elaborated in Section IV, including system throughput and idle codebook probability. Subsequently in Section V, with the assist of the derived analytical results, a user barring mechanism is designed. Then, we discuss the practical application of our proposed scheme in Section VI. Simulation results are provided in Section VII to validate the derived theoretical results and proposed user barring scheme. Finally, Section VIII concludes the paper.

\emph{Notations:} Scalar variables are denoted by normal-face letters, while boldface capital and blodface lowercase symbols represent matrices and column vectors, respectively. ${( \cdot)^T}$ operation stands for matrix transpose. In addition, $\odot$ denotes the Hadamard product. $\text{diag}(\cdot)$ denotes a diagonal matrix with the elements on its diagonal. ${\tbinom{n}{k}} $ and $\left\lceil  \cdot  \right\rceil $ refer to the binomial coefficient and the ceil operation, respectively.

\begin{figure*}[!t]
\centering
\includegraphics[width=135mm]{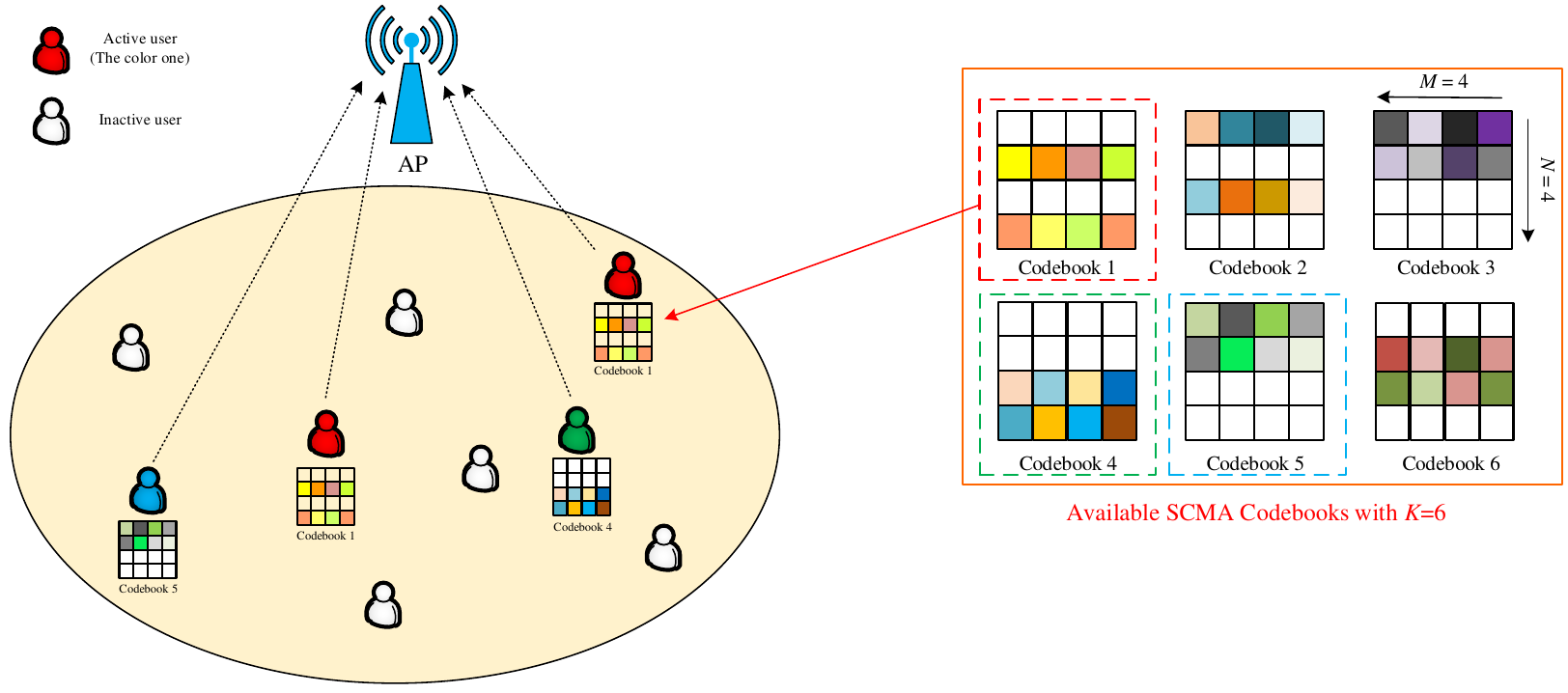}
\caption{Illustration of the proposed SCMA-empowered URA system, where an AP serves for numerous users and the scale of the available SCMA codebook pool is $N =4$, $K=6$ and $M = 4$. In a transmission slot, each active user randomly selects an SCMA codebook from the available codebook set to convey data information. As shown, users marked by red, green and blue choose codebook 1, 4 and 5, respectively. Particularly, it can be observed that two active users concurrently selects codebook 1, resulting in SCMA codebook collisions, which is inevitable in the proposed URA system. This phenomenon arises from the lack of central coordination, which is able to allocate an individual SCMA codebook to each user and avoid collisions.}
\end{figure*}

\section{Preliminaries and System Model}
\subsection{SCMA}
In SCMA, users' information bits are spread over the same $N$ orthogonal frequency resources. Specifically, each active user utilizes a pre-defined SCMA codebook and the SCMA encoder directly maps its $\log_2(M)$ data bits to an $N$-dimensional complex sparse codeword. There are $K$ available codebooks in an SCMA system and the overloading factor $\theta$ is defined as $ \theta = \frac{K}{N}> 100\%$. Moreover, $K$ can be also explained as the maximum number of active users an SCMA system can accommodate simultaneously.  An arbitrary SCMA codebook $k$ contains a total of $M$ SCMA codewords, denoted by $\mathcal {X}_{k} \in {{\mathbb{C}}^{N \times M}}$, which is given as $\mathcal {X}_{k}=\{\mathbf {x}_{k1},\ldots \mathbf {x}_{km},\ldots,\mathbf {x}_{kM}\}$. Particularly, $M$ denotes the cardinality, which is determined by the modulation order. Each SCMA codeword is a sparse vector and only a few components inside are non-zero elements. Furthermore, codewords from different codebooks differ in at least one non-zero position, while codewords from the same codebook hold common non-zero positions. Owing to the dedicated codebook design, the non-zero elements belonging to the same codebook are individual among each other, thus facilitating the decoding process.

To show the sparse feature of SCMA codewords explicitly, a sparse matrix ${\bf{F}} \in {{\mathbb{C}}^{N \times K}}$, namely an indicator matrix, is introduced. The element in ${\bf{F}}$ is defined as ${f_{n,k}}$.
${{\bf{f}}_k} \in {{\mathbb{C}}^{N \times 1}}$ denotes the indicator vector for codebook $k$, which can be expressed as ${{\bf{f}}_k} = \left( {{f_{1,k}},{f_{2,k}},...,{f_{N,k}}} \right)^T$. Accordingly, we have ${\bf{F}} = \left( {{{\bf{f}}_1},{{\bf{f}}_1},...,{{\bf{f}}_K}} \right)$. A concrete example ${{\bf{F}}_{4 \times 6}}$ is given in \eqref{system_1}, where $N=4$, $K=6$ and $\theta= 150{\rm{\% }}$. For ease of notation, an available frequency resource is defined as a frequency node (FN). In ${\bf{F}}$, the set of nonzero elements in each row corresponds to SCMA codebooks which occupy the same FN while the ones in each column represent the set of FNs a codebook adopts to transmit data information. Codebook $k$ connects with FN $n$ if and only if ${f_{n,k}}=1$. 

\begin{equation}\label{system_1}
\small
{{\bf{F}}_{4 \times 6}} = \left[ {\begin{array}{*{20}{c}}
0&1&1&0&1&0\\
1&0&1&0&0&1\\
0&1&0&1&0&1\\
1&0&0&1&1&0
\end{array}} \right].
\end{equation}

Let ${{\xi }_{n}}= \left\{ {k\left| {{f_{n,k}} \ne 0} \right.} \right\}$ and ${{\zeta }_{k}}=\left\{ {n\left| {{f_{n,k}} \ne 0} \right.} \right\}$ are the set of codebooks sharing the $n$th FN and the set of FNs occupied by codebook $k$, respectively. In addition, the number of codebooks which share the same FNs is defined as $d_f$, i.e. $\left| {{\xi }_n} \right| = {d_f}$, and the number of FNs occupied by codebook $k$ is named as $d_v$, i.e. $\left| {{{\zeta }_k}} \right| = {d_v}$. Hence, in \eqref{system_1}, it can be seen that $d_v = 2$ and $d_f = 3$.


\subsection{Proposed SCMA-empowered S-ALOHA Scheme}
In this subsection, a novel URA scheme with SCMA is introduced, which is shown in Fig. 1. Considering a single cell communication system, an AP serves a massive number of IoT users. The S-ALOHA scheme is applied to uplink access, where the time is divided into several transmission slots. Before each transmission slot, the AP firstly broadcasts beacon signals to inform active IoT users to send out their packets in certain transmission slots. To be specific, beacon signals are not only used to carry out uplink synchronization, but also convey SCMA information for the following uplink access. Moreover, each active IoT user is capable of estimating its own channel state information (CSI) based on the received beacon signal, owing to the channel reciprocity in time division duplexing (TDD) mode \cite{NOSA-1}. In this paper, we assume that the uplink and downlink channel states would not vary rapidly and remain unchanged for a period of time.

In every transmission slot, each active IoT user randomly selects an SCMA codebook out of the available codebook pool with equal probability to deliver data packet. Assume that user $i$ is active and selects codebook $k$ at the $t$th time slot. After mapping information bits according to the selected codebook, user $i$'s complex SCMA codeword ${{\bf{x}}_{i,t}} \in {{\mathbb{C}}^{N \times 1}}$ coming from codebook $k$ is transmitted to the AP. The received superimposed signal ${{\bf{y}}_t} \in {{\mathbb{C}}^{N \times 1}}$ at the AP is expressed as
\begin{equation}\label{system_2}
{{\bf{y}}_t} = \sum\limits_{i \in {\mathcal{I}_{t}}} {{\text{diag}}\left( {{{\bf{h}}_{i,t}} \odot {{\bf{w}}_{i,t}}} \right){{\bf{x}}_{i,t}}}  + {\bf{z}},
\end{equation}
where $\mathcal{I}_{t}$ stands for the set of active users at the $t$th time slot. ${{\bf{h}}_{i,t}} \in {{\mathbb{C}}^{N \times 1}}$ denotes the channel coefficient of user $i$ at the $t$th time slot. In particular, due to the sparse nature of SCMA codewords, ${{\bf{h}}_{i,t}}$ is a sparse vector with $d_v$ non-zero elements. ${{{\bf{w}}_{i,t}}} \in {{\mathbb{C}}^{N \times 1}} $ is the pre-equalization vector of user $i$ for the channel effect elimination. ${{{\bf{w}}_{i,t}}}$ is designed based on the following rules:
\begin{equation}\label{system_3}
{{{\bf{h}}_{i,t}} \odot {{\bf{w}}_{i,t}}} = {{\bf{f}}_k}.
\end{equation}
Recall that the channel information can be estimated by the beacon signal. In this paper, the perfect CSI estimation is assumed. Moreover, ${\bf{z}} \in {{\mathbb{C}}^{N \times 1}}$ represents the complex additive white Gaussian noise (AWGN) vector with zero mean and $N_0{{\bf{I}}_N}$ variance.

In the proposed SCMA-empowered URA scheme, due to the lack of central coordination and interactions among active users, codebook collisions unavoidably take place, especially when the system becomes congested. In particular, codebook collision indicates that multiple active users select the same SCMA codebook for the uplink data transmission. In the existing literature \cite{SCMA-3}-\cite{SCMA-7} which focused on the uplink SCMA systems, the AP is assumed to allocate each active user with an individual SCMA codebook. In this case, the presence of codebook collisions is ignored, which is the major difference between the scheduled access scheme and the proposed URA-based scheme. Here, we use $a_{k,t}$ to denote the number of active users selecting codebook $k$ at $t$th time slot. There are three possible values for $a_{k,t}$, given as
\begin{equation}\label{system_4}
{a_{k,t}} = \left\{ {\begin{array}{*{20}{l}}
0,&{{\rm{selected \; by \; none \; user}}}\\
1,&{{\rm{selected \; by \; a \; single \; user}}}\\
{{\delta _k}},&{{\rm{selected \; by \; more \; than \; one \; user}}}
\end{array}} \right.,
\end{equation}
where ${{\delta _k}} > 1$. The pattern of the selected codebooks is written as ${\bf{a}}_t = {[{{a_{1,t}},{a_{2,t}},...,{a_{K,t}}} ]^T}$. Furthermore, we introduce an active pattern indicator matrix, termed as ${\bf{A}}_{t} \in {{\mathbb{C}}^{N \times K}}$, to intuitively indicate the state of available SCMA codebooks at the $t$th time slot. $\bf{A}_{t}$ is generated from the indicator matrix ${\bf{F}}$, expressed by
\begin{equation}\label{system_5}
{{\bf{A}}_t} = \text{diag}({{\bf{a}}_t}){\bf{F}},
\end{equation}
where its $k$th column represents the number of active users choosing codebook $k$ and $n$th row denotes the number of active users concurrently share FN $n$. For instance, when the SCMA codebook indicated by \eqref{system_1} is employed and assume ${\bf{a}}_t = {[{{\delta _1},1,0,0,0,1} ]^T}$, ${{\bf{A}}_t}$ is given as
\begin{equation}\label{system_6}
\small
{\bf{A}}_t = \left[ {\begin{array}{*{20}{c}}
0&1&0&0&0&0\\
{\delta _1}&0&0&0&0&1\\
0&1&0&0&0&1\\
{\delta _1}&0&0&0&0&0
\end{array}} \right],
\end{equation}
which indicates that a codebook collision happens at the first SCMA codebook. Two codebooks are selected by a single user and no one chooses the third, fourth and fifth codebooks. Besides, in terms of the frequency resource, only one user occupies the first FN while several users share the second FN at the same time. Noting that, when ${\bf{a}}_t = {[{1,1,...,1} ]^T}$, ${{\bf{A}}_t} = {\bf{F}}$, indicating that the conventional uplink SCMA system investigated in \cite{SCMA1-3} is a special case in the proposed SCMA-aided S-ALOHA scheme. Without loss of generality and for simplicity, we omit the subscript $t$ in the sequel.

\section{Proposed IC-first Blind Decoding Strategy }
Owing to the sparsity of SCMA codewords, MPA can be utilized at the receiver side to achieve maximum likelihood performance with relatively low computational complexity. The conventional MPA algorithm is widely used in the scenario where ${\bf{a}} = {[ {1,1,...,1}]^T}$ is assumed. Specifically, the issue of user activity detection is neglected and each user is assumed to hold an individual SCMA codebook. In mMTC scenarios, taking the user sparsity into consideration, the JMPA detector proposed by \cite{SCMA-3} is capable of blindly decoding the users’ data with no knowledge of user activity state by searching all the possible combinations of active users. However, JMPA targets at the contention-free grant-free RA, where each user is pre-allocated a unique SCMA codebook. More precisely, JMPA exclusively serves for the case where $a_k = 0$ or $a_k = 1$ in \eqref{system_5}. Conversely, in the proposed SCMA-empowered S-ALOHA scheme, each active user follows arrive-and-go manner, where it can randomly select an SCMA codebook to transmit without scheduling, thus leading to inevitable codebook collisions. Under this circumstance, JMPA method may fail to work with the appearance of codebook collisions, resulting in the failure of the uplink access. Thus, JMPA method cannot be directly applied to our proposed URA system. 

\begin{figure*}[!t]
\centering
\includegraphics[width=145mm]{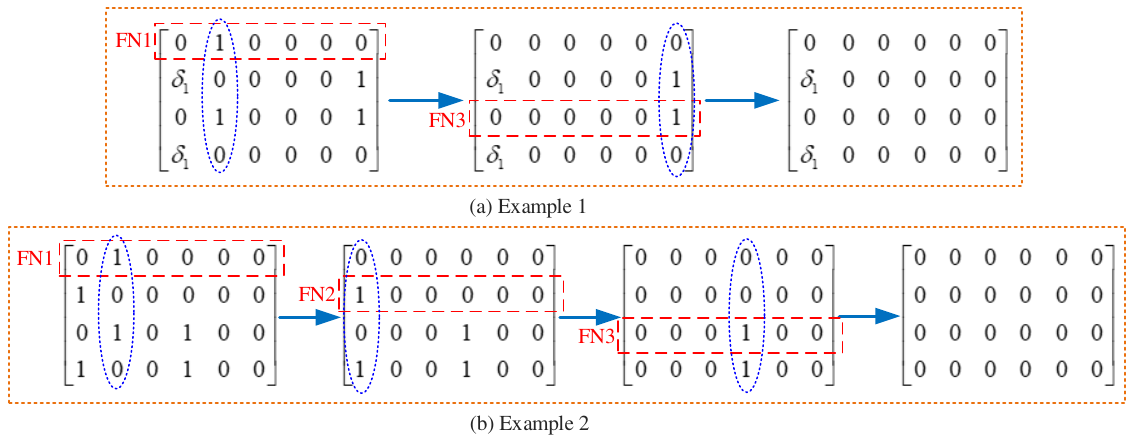}
\caption{The schematic diagram of decoding process by the proposed IC-first blind decoding strategy.}
\label{flow_diagram}
\vspace{-1em}
\end{figure*}

To address this challenge, an efficient blind decoding strategy is developed tailored for the scenarios where codebook collisions occur, which integrates IC and JMPA algorithms, called as IC-first decoding strategy. In the proposed decoding strategy, the first step is to carry out IC sequentially based on the index of FNs, targeting at removing the inter-user interferences to the greatest extent, which skillfully utilize the sparse structure of SCMA codebooks. It is noted that the core idea of IC is to find out FNs without inter-user interference, namely single FN. Considering the orthogonality among FNs and inter-user interference-free, this part of information is decodable. Moreover, when one of the non-zero component in an SCMA codeword has been successfully decoded, the whole SCMA codeword is accordingly obtained thanks to its specific multidimensional nature. Then, the contribution of the decoded codewords in the received superimposed signal can be removed. Subsequently, the AP continues to find single FN. Until there is no single FN, the JMPA algorithm is performed repeatedly to blindly decoding the remaining superimposed signals, which cannot be canceled by the former IC. Table I presents the applicable conditions for various decoding strategies based on the possible values of $a_k$ in \eqref{system_4}.

{\renewcommand\arraystretch{1.5}
\begin{table}
\footnotesize
\centering
{\caption{ Applicable scenarios for Various Decoding Methods}}
{\begin{tabular}{|l|l|l|}\hline
Detection Methods & Value of $a_k$ & Scenarios\\\hline
Conventional MPA \cite{NOMA-3}-\cite{SCMA1-3} &  1 &  Without collisions\\\hline
JMPA \cite{SCMA-3}-\cite{SCMA-6} & 0 or 1 &  Without collisions\\\hline
Proposed strategy & 0, 1, or $\delta$ & Collisions occur\\\hline
\end{tabular}}
\end{table}}

The advantages of the proposed decoding strategy mainly manifest in two aspects: high system throughput and low computational overhead, respectively. To be specific, high system throughput means that the proposed decoding strategy enables to decode part of transmitted data information even in the presence of codebook collisions, owing to the exploitation of the IC technique. Additionally, in certain cases, IC method can completely replace the MPA-based method to complete the whole decoding process with a low computational cost. For ease of understanding, we provide two concrete cases to illustrate the above two advantages. Moreover, a schematic diagram is given in Fig. \ref{flow_diagram} to vividly exhibit the decoding process of the two examples using our proposed decoding strategy.

\textit{\textbf {Example 1}: Take the active pattern indicator matrix given in \eqref{system_6} as an example. In this case, it is apparent that a codebook collision happens at the first codebook, thus the JMPA detector loses its effectiveness at the receiver. However, according to the principle of the proposed IC-first decoding strategy, packets using second and sixth codebooks are decodable. The decoding procedure can be described as follows, which is also shown in  Fig. \ref{flow_diagram}(a). In the beginning, the AP is able to observe that the first frequency resource is only used by one user, thus regarding it as a single FN. The packet occupying the first FN with the second codecook is decodable and can be removed from the received signals. After performing cancellation, the third FN switches as a new single FN. Correspondingly, the packet with the sixth codebook is obtained. Then, there is no single FNs and JMPA method is employed to continue the decoding process. Due to the appearance of codebook collisions, JMPA method fails to work and the decoding process gets stuck. To sum up, under \eqref{system_6} case, there are total four active users attempting to access. The proposed decoding strategy is able to identify two successful accesses, while that of using the existing JMPA method is zero.}

\textit{\textbf{Example 2:} To explain the advantage of low computational overhead, we assume a case with ${\bf{a}} = {\left( {1,1,0,1,0,0} \right)^T}$ and the corresponding active pattern matrix is given as
\begin{equation}\label{system_7}
\small
{\bf{A}}^{(1)} = \left[ {\begin{array}{*{20}{c}}
0&1&0&0&0&0\\
1&0&0&0&0&0\\
0&1&0&1&0&0\\
1&0&0&1&0&0
\end{array}} \right].
\end{equation}
Since there is no codebook collision in ${\bf{A}}^{(1)}$, JMPA method is available but with the computational complexity $\mathcal O\left( {{{\left( {M + 1} \right)}^{{d_f}}}} \right)$. Conversely, when adopting the proposed IC-first framework, the received signal can be successfully decoded by IC technique only with the decoding order of the second, fourth and first codebooks. The concrete decoding process can be found in Fig. \ref{flow_diagram}(b). Then, the decoding process will terminate when the IC process finishes. Under this case, the complexity of the decoding process is $\mathcal O\left({M}{d_f}\right)$, which is negligible compared to that of utilizing JMPA.
}

\section{Theoretical analysis of the proposed scheme}
In this section, the theoretical analysis of our proposed scheme is conducted. Firstly, we investigate the system throughput and derive a theoretical closed-form expression. Subsequently, the idle codebook probability is studied and the corresponding theoretical expression is given, which enables the real-time traffic load estimation. During our theoretical analysis, it is assumed that the number of active IoT users at each transmission time slot follows the Poisson distribution with arrival rate $\overline \lambda$. Hence, the number of active IoT users randomly choosing an arbitrary SCMA codebook follows the Poisson distribution with mean $\lambda  = \frac{{\overline \lambda  }}{K}$. Accordingly, the probability of total $j$ active users concurrently choosing one codebook can be expressed as ${p_j} = \frac{{{\lambda ^j}{e^{ - \lambda }}}}{{j!}}$. Then, the probabilities that an arbitrary SCMA codebook is chosen by only a single active user or be selected by no one are denoted as ${p_1} = \lambda {e^{ - \lambda }}$ and ${p_0} = {e^{ - \lambda }}$, respectively.

\subsection{System Throughput}
In this subsection, we systematically analyze the system throughput by leveraging the unique codebook properties of SCMA. Note that the given theoretical analysis fully incorporates our proposed IC-first blind decoding capability, particularly tailored for collision scenarios.

For an active IoT user $i$, its transmitted SCMA codeword can be successfully decoded by the AP when satisfying at least one of the following conditions:
\begin{itemize}
\item [({a})]
{ There is no codebook collision, i.e., each active IoT user selects a unique SCMA codebook, like the situation in \eqref{system_7}.
}
\end{itemize}

\begin{itemize}
\item [({b})]
{ Codebook collisions exist, but the codebook selected by user $i$ occupies a FN without inter-user interference, i.e., there exists a single FN for user $i$. Thus, the user $i$'s SCMA codeword can be decoded by IC, like the user with the second codebook in \eqref{system_6}.
}
\end{itemize}

\begin{itemize}
\item [({c})]
{There exist codebook collisions and there is no single FNs for user $i$. However, as the IC process proceeds, part of the inter-user interference can be removed and there will be a new single FN for user $i$, like the user with the sixth codebook in \eqref{system_6}.
}
\end{itemize}
For user $i$, the aforementioned three conditions are independent and it is impossible for any two of them to simultaneously appear. In the sequel, we analyze the system throughput separately based on one of them. Without loss of generality, we consider the case with $d_v = 2$, i.e., each SCMA codeword occupies $d_v = 2$ FNs and the remaining $N-d_v$ elements are zeros, which is a common case in the state-of-the-art researches \cite{NOMA-3}, \cite{SCMA1-4}, \cite{SCMA-6}. It should be also noted that the following theoretical results can be easily extended to the cases with $d_v > 2$ when taking more possible situations into account. For ease of understanding, we define the index of FNs occupied by codebook $k$ as $x$ and $y$, i.e., ${{\zeta }_{k}} = \left\{ x, y \right\}$.

\textit{\textbf {Case 1}: Condition (a) is met.} 
Under this case, the probability of user $i$ randomly selecting one SCMA codebook without collisions is $p_1$. Furthermore, the probability that no codebook collision appears at the remaining $K-1$ codebooks is given as $(p_1 + p_0)^{K-1}$. Therefore, considering total $K$ available SCMA codebooks, the throughput of meeting condition (a) can be expressed as
\begin{equation}\label{SCMA_1}
{T_1} = K{p_1}{\left( {{p_0} + {p_1}} \right)^{K - 1}}.
\end{equation}
It should be remarked that if the existing JMPA method is utilized to cope with the decoding problem in the proposed SCMA-aided S-ALOHA system, \eqref{SCMA_1} is the exact analytical expression for the system throughput.

\textit{\textbf {Case 2}: Condition (b) is met.}  
Assume that codebook $k$ is solely selected by user $i$. Following condition (b), at least one codewords out of $d_v$ non-zero codewords in the codebook $k$ occupies one FN without inter-user interference. We focus on the situation that FN $x$ occupied by one of the non-zero codewords is interference-free, i.e., FN $x$ is a single FN, only occupied by user $i$. Thus, codebooks occupying FN $x$ cannot be used, except codebook $k$. The corresponding probability can be given as ${p_1}{({{p_0}})^{{d_f} - 1}}$. Next, considering codebook collisions, it is noted that collisions must occur at the codebooks which hold zero element at FN $x$ with the corresponding probability as ${1 - {{( {{p_0} + {p_1}})}^{K - {d_f}}}}$. Consequently, since FN $x$ and $y$ are independent and there are total $K$ available SCMA codebooks, the throughput of this case can be computed by the following equation:
\begin{equation}\label{SCMA_2}
\begin{aligned}
{T_2} &= K{\tbinom{d_v}{1}}{p_1}{\left( {{p_0}} \right)^{{d_f} - 1}}\left( {1 - {{\left( {{p_0} + {p_1}} \right)}^{K - {d_f}}}} \right)\\
&=2K{p_1}{\left( {{p_0}} \right)^{{d_f} - 1}}\left({1 - {{\left( {{p_0} + {p_1}} \right)}^{K - {d_f}}}} \right).
\end{aligned}
\end{equation}
It is crucial to remark that the cases that both two non-zero codewords in the codebook $k$ occupy single FNs separately, i.e., both FN $x$ and $y$ are single FNs, has been included in \eqref{SCMA_2}.

\textit{\textbf {Case 3}: Condition (c) is met.} Under this circumstance, the potential cases will increase with the size of the SCMA codebook. For simplicity, we consider a most possible situation that collisions occur at FN $y$ and FN $x$ are concurrently occupied by two active users, one selecting codebook $k$ and another one selecting codebook $\overline k$. Thus, even codebook $k$ is selected by only one user, the corresponding information cannot be directly decoded. Fortunately, the user choosing codebook $\overline k$ holds a single FN $z$, thus its information is able to be removed by performing IC. In this case, the probability that one user choosing codebook $\overline k$ shares FN $x$ with the one choosing codebook $k$ is ${p_1}\left( {\left( {{d_f} - 1} \right){{\left( {{p_0}} \right)}^{{d_f} - 2}}{p_1}} \right)$. Moreover, the case that FN $z$ is single indicates that codebooks occupying FN $z$ cannot be selected, except codebook $\overline k$, whose probability is expressed as ${({{p_0}})^{{d_f} - 1}}$. It should be noted that no one selects the codebook occupying FN $y$ and $z$ and the corresponding probability has been included in the above expression. Finally, considering the probability of codebook collisions, we can have a lower bound throughput for the cases meeting condition (c) as follows:
\begin{equation}\label{SCMA_3}
\begin{aligned}
{T_3} &\approx 2K{p_1}\left( {\left( {{d_f} - 1} \right){{\left( {{p_0}} \right)}^{{d_f} - 2}}{p_1}} \right){\left( {{p_0}} \right)^{{d_f} - 1}}\\
&\times \left( {1 - {{\left( {{p_0} + {p_1}} \right)}^{{d_f} - 2}}} \right).
\end{aligned}
\end{equation}
As a consequent, the throughput of the proposed SCMA-empowered URA scheme can be calculated by summing up \eqref{SCMA_1}, \eqref{SCMA_2} and \eqref{SCMA_3} as
\begin{equation}\label{SCMA_4}
T = {T_1} + {T_2} + {T_3}.
\end{equation}
Noting that since a few potential cases are discarded when counting the system throughput under condition (c), the analytical expression of the system throughput in \eqref{SCMA_4} is slightly less than the actual throughput, especially with the increasing SCMA codebook size.

Thanks to the inherent attribute of SCMA, which allows for accommodating a larger number of active users compared to available FNs, the maximum throughput of the proposed SCMA-empowered URA system enjoys a notable enhancement when compared to the OMA-based S-ALOHA system. Notably, the theoretical optimal traffic load $\lambda ^ *$ corresponding to the maximum system throughput can be obtained based on \eqref{SCMA_4}. Note that when the actual traffic load is below $\lambda ^ *$, the throughput exhibits an upward trend with the traffic load, since the system doesn't reach its saturation point, referred as underloading state. However, once the actual traffic load surpasses $\lambda ^ *$, the system throughput undergoes a decline from the peak due to increased collisions, referred as overloading state. Particularly in the context of the SCMA-empowered URA system, given that each active user shares multiple FNs with others, collisions at each FN become severe, thus resulting in a sharper throughput degradation compared to that of OMA-based URA systems, especially under heavy traffic loads. To address this challenge, we adopt a user barring method to regulate the real-time traffic load, aiming to reduce collisions and keep the throughput at the peak.

\subsection{Idle SCMA codebook probability}
In order to implement the user barring mechanism, a reliable user load estimation is necessary. In this paper, a novel parameter, namely idle SCMA codebook probability $P_\text{idle}$, is employed, facilitating the estimation of the real-time traffic load. For SCMA codebook $k$, it can be regarded as an idle codebook at a certain time slot only if the AP can observe that no one uses codebook $k$. It needs to be emphasized that the AP may lack the ability to detect the state of some SCMA codebooks during the decoding process, unless there is no codebook collision. In other words, although some codebooks are unoccupied, the AP is blind to their states. Here, we give two concrete examples to explicitly illustrate this point.



\begin{equation}\label{SCMA_22}
\small
{\bf{A}}^{(2)} = \left[ {\begin{array}{*{20}{c}}
0&{\delta _2}&0&0&0&0\\
{\delta _1}&0&0&0&0&0\\
0&{\delta _2}&0&0&0&0\\
{\delta _1}&0&0&0&0&0
\end{array}} \right].
\end{equation}

\begin{equation}\label{SCMA_33}
\small
{\bf{A}}^{(3)} = \left[ {\begin{array}{*{20}{c}}
0&{\delta _2}&{\delta _3}&0&0&0\\
0&0&{\delta _3}&0&0&0\\
0&{\delta _2}&0&0&0&0\\
0&0&0&0&0&0
\end{array}} \right].
\end{equation}
Note that the above \eqref{SCMA_22} and \eqref{SCMA_33} are generated from \eqref{system_1} according to \eqref{system_5}. In \eqref{SCMA_22}, even though four codebooks are unused, the AP fails to detect them since codebook collisions happening at the first and second codebook causes all FNs unobservable. On the other hand, when decoding the case with \eqref{SCMA_33}, the AP can find that no one occupies the fourth FN, thus it is able to know that codebooks sharing the fourth FN are idle, i.e., the first, fouth and fifth codebooks are unused according to \eqref{system_1}. Despite of this, the AP still fails to determine that total four codebooks are unselected in \eqref{SCMA_33}. Particularly, the AP is blind to the state of the sixth codebook, since the FNs its non-zero codewords separately occupy suffer from collisions and become undecodable. To sum up, counting the number of idle codebooks by the AP heavily depends on the decoding results, which is influenced by the distribution of active users' selection. Based on the distribution of active users, we investigate the idle SCMA codebook probability in different cases. To ease the representation of the following analysis, we introduce two auxiliary definitions.

The first one is called as orthogonal SCMA codebooks. For codebook $k$, its orthogonal SCMA codebooks must occupy different FNs compared with that of codebook $k$, like the first and second codebooks in \eqref{system_1}. The number of orthogonal SCMA codebooks for an arbitrary codebook $k$ equals to
\begin{equation}\label{SCMA_5}
N_\text{orth} = {\tbinom{N-d_v}{d_v}}.
\end{equation}


Next, recall that each FN connects with total $d_f$ SCMA codebooks. For an arbitrary FN $n$, assuming that no one selects the codebook $k$ connecting with FN $n$, the case that none of the remaining $d_f-1$ codebooks connecting with FN $n$ are selected or at most one of them is selected is defined as $\Psi$. When $\Psi$ appears, the state of FN $n$ can be observed by the AP, idle or single, respectively. Then, according to the definition, the probability that $\Psi$ happens is termed as $\phi$, given as
\begin{equation}\label{SCMA_6}
\phi  = {( {{p_0}})^{{d_f} - 1}} + ( {{d_f} - 1} ){({{p_0}})^{{d_f} - 2}}{p_1}.
\end{equation}

The first situation corresponds to no codebook collisions. Hence, once no one selects SCMA codebook $k$, it can be observed by the AP and regarded as an idle codebook. The probability of this case is
\begin{equation}\label{SCMA_7}
{P_\text{idle}^{(1)}} = {p_0}{\left( {{p_0} + {p_1}} \right)^{K - 1}}.
\end{equation}

Next, we consider the scenarios with the appearance of codebook collisions. However, when codebook collisions occur, the analysis of the idle codebook probability becomes complicated. In this case, codebook $k$ is regarded as an idle codebook if and only if at least one of FNs that codebook $k$ occupies must be observed by the AP. Like the derivation in Section IV, the index of FNs occupied by codebook $k$ is still defined as $x$ and $y$. Here, we first study the case when one of FNs can be observed. Since $x$ and $y$ are independent and inter-changable, the probability can be computed by $2{p_0}\phi$. Furthermore, two cases need to be separately discussed. The first one is that FN $y$ exists codebook collisions, while the second one is that codebook collisions only take place at the orthogonal codebooks of codebook $k$. Focusing on the first case, the corresponding probability can be computed by ${1 - {{( {{p_0} + {p_1}})}^{{d_f} - 1}}}$. Therefore, we have
\begin{equation}\label{SCMA_8}
{P_\text{idle}^{(2)}} = 2{p_0}\phi\left({1 - {{( {{p_0} + {p_1}})}^{{d_f} - 1}}}\right).
\end{equation}
In addition, the second case indicates that collisions happen at the orthogonal codebooks and there is no collision at FN $y$. Therefore, we can derive the probability of the second case as ${({{p_0} + {p_1}})^{{d_f} - 1}}\left({1 - {{( {{p_0} + {p_1}})}^{{N_{\text{orth}}}}}} \right)$. The third part of the idle codebook probability can be represented by
\begin{equation}\label{SCMA_9}
{P_\text{idle}^{(3)}} = 2{p_0}\phi{({{p_0} + {p_1}})^{{d_f} - 1}}\left({1 - {{( {{p_0} + {p_1}})}^{{N_{\text{orth}}}}}} \right).
\end{equation}
However, one case has been repetitively counted by \eqref{SCMA_8} and \eqref{SCMA_9}. More precisely, FN $x$ and $y$ are both observable and collisions happen at the orthogonal codebooks. The reason is that when investigating FN $x$ or $y$, the state of the other one is ignored. The probability that FN $x$ and $y$ are both observable is written as ${p_0}{\phi}^2$. In addition, the probability that collisions happens at the orthogonal codebooks has been computed in \eqref{SCMA_9}, given as $\left({1 - {{( {{p_0} + {p_1}})}^{{N_{\text{orth}}}}}} \right)$. Hence, the probability of the repetitive part can be expressed by
\begin{equation}\label{SCMA_10}
{P_\text{idle}^{(4)}} = {p_0}{{\phi}^2}\left({1 - {{( {{p_0} + {p_1}})}^{{N_{\text{orth}}}}}} \right).
\end{equation}
As a result, combining \eqref{SCMA_7}-\eqref{SCMA_10}, the idle SCMA codebook probability can be calculated by \eqref{SCMA_11}, as shown.
\begin{equation}\label{SCMA_11}
{P_\text{idle}} = {P_\text{idle}^{(1)}} + {P_\text{idle}^{(2)}} + {P_\text{idle}^{(3)}} - {P_\text{idle}^{(4)}}.
\end{equation}
By substituting \eqref{SCMA_5}, \eqref{SCMA_6}, $p_0$ and $p_1$ into \eqref{SCMA_11}, the exact theoretical expression of ${P_\text{idle}}$ can be obtained, which is a univariate function in terms of $\lambda$. Intuitively, when the traffic load intensifies, i.e., $\lambda$ increases, the number of idle codebooks decreases. This implies that ${P_\text{idle}}$ exhibits a monotonic descent with respect to $\lambda$, which is further proved by our simulation results. In other words, a distinct idle codebook probability corresponds to a specific traffic load, which contributes to the load estimation.

\section{User Barring Mechanism for the proposed URA Schemes}
As aforementioned, when a great number of IoT users attempt to access simultaneously, codebook collisions become exacerbated, subsequently resulting in a substantial degradation of the system throughput. In order to mitigate the detrimental impact stemming from the traffic congestion, a user barring mechanism is designed for the proposed SCMA-aided S-ALOHA system, which can effectively manage the traffic load and regulate the number of users concurrently seeking access the system. Its objective is to ensure that the actual access load remains close to the optimal one, thereby enabling the system to achieve the maximum throughput.

\begin{figure}[t]
\centering{\includegraphics[width=60mm]{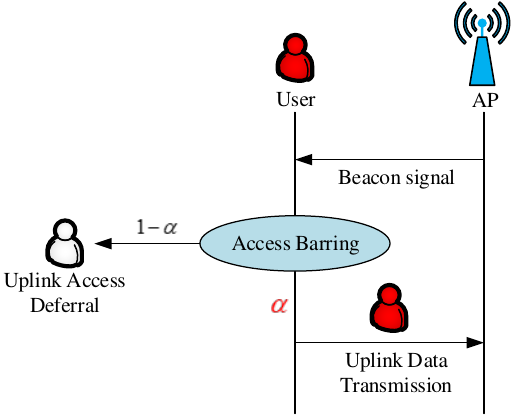}}
\caption{The diagram of the implementation of user barring mechanism.}
\label{barring_pic}
\vspace{-1em}
\end{figure}

Prior to the transmission phase, the AP broadcasts an access probability $\alpha$, also denoted as barring rate, to all users through the beacon signal. Then, users seeking uplink access transmit their packets with probability $\alpha$. Specifically, each active user randomly produces a value between 0 and 1 before the uplink transmission. If the value is less than or equal to $\alpha$, the user implements the uplink access. Otherwise, its access attempt has to be deferred. Fig. \ref{barring_pic} exhibits the implementation process of the user barring mechanism. In this paper, we assume that each frame contains a successive $J$ transmission slots and the AP broadcasts $\alpha$ at the beginning of each transmission frame. It indicates that the access probability $\alpha$ remains constant throughout an entire frame duration and is updated at the beginning of the subsequent frame.

\begin{algorithm}[t]
\caption{User Barring Algorithm}
\label{algorithm 1}
\begin{algorithmic}[1]
\REQUIRE ${\lambda ^ * }$, ${P_\text{idle}^*}$, ${{\alpha}_\text{old}}$, $\Omega$, $J$;\\
Repeat the following procedures before each transmission frame.
\STATE Count the average throughput $\widetilde T$.

\STATE Obtain two possible loads ${\lambda_\text{under}}$ and ${\lambda_\text{over}}$ by referring the pre-determined load-throughput table $\Omega$.

\STATE Count the average idle codebook probability ${\tilde {P}}_\text{idle}$.

\STATE Obtain the estimated load $\hat {\lambda}$ by comparing ${\tilde {P}}_\text{idle}$ with ${P_\text{idle}^*}$.

\STATE Update the access probability ${{\alpha}_\text{new}}$ according to \eqref{SCMA_12}.

\ENSURE ${{\alpha}_\text{new}}$.\\

\end{algorithmic}

\end{algorithm}

Within the user barring mechanism, an effective traffic load control is achieved by dynamically adjusting the access probability $\alpha$ based on the real-time user traffic load. Hence, a reliable approach to estimate the load during each transmission frame is imperative. In this paper, the derived analytical expressions of the system throughput \eqref{SCMA_4} and idle codebook probability \eqref{SCMA_11} are utilized to estimate the actual traffic load. In particular, a pre-determined load-throughput table $\Omega$ can be obtained by \eqref{SCMA_4}, which stores different throughput values alongside their corresponding load values \cite{NOSA-3}. Such a table provides a valuable reference for the load estimation based on the observed throughput during each transmission frame. Concretely, once the AP decodes the received signal, an average throughput $\widetilde T$ per time slot in the current frame can be counted. By referencing the table and comparing the observed throughput to the corresponding load values, an accurate estimation of the current user load can be obtained. However, two distinct load values can be found, referred to as ${\lambda_\text{under}}$ in the underloading state and ${\lambda_\text{over}}$ in the overloading state, respectively. This is because the system throughput presents earlier increase and later decrease trend with the increasing traffic load. Fortunately, owing to the monotonic property of the idle codebook probability, the estimated load can be uniquely determined. To be specific, based on the optimal load $\lambda ^ *$, ${P_\text{idle}^*}$ is accordingly obtained. After each transmission frame, an average idle codebook probability ${\tilde {P}}_\text{idle}$ can be observed by the AP. If ${\tilde {P}}_\text{idle} >{P_\text{idle}^*}$, the system is judged as in the underloading state and ${\lambda_\text{under}}$ is regarded as the estimated load $\hat {\lambda}$. Conversely, ${\tilde {P}}_\text{idle} \le  {P_\text{idle}^*}$ indicates the overloading state and $ \hat {\lambda} = {\lambda_\text{over}}$.

After obtaining the estimated load $\hat {\lambda}$, the newly access probability ${{\alpha}_\text{new}}$ for the next transmission frame can be given as
\begin{equation}\label{SCMA_12}
{{\alpha}_\text{new}}  = \min \left( {1,{{\alpha}_\text{old}} \frac{\lambda ^ * } {\hat {\lambda}}} \right),
\end{equation}
where ${{\alpha}_\text{old}}$ denotes the access probability applied to the last transmission frame. The procedure of the user barring mechanism in the SCMA-aided S-ALOHA scheme is detailed in \textbf{Algorithm 1}.

\section{Discussion}
In this section, with the aim of enhancing the practical applicability of our proposed SCMA-empowered SCMA scheme, we discuss two popular use cases, i.e., mMTC and satellite-terrestrial communication, respectively.
\subsection{The application within mMTC scenarios}
In the context of mMTC, it is evident that the AP needs more physical resources to accommodate massive users. We assume that there are $N_{\text{total}}$ available frequency resources. The first approach is to design a large-scale SCMA codebook along with the $N_{\text{total}}$ available frequency resources. However, it significantly incurs the increasing complexity of codebook design as well as the decoding process. Alternatively,  we devise a group RA strategy to cope with the challenges  when encountering massive access. To be specific,  $N_{total}$ available frequency resources are partitioned into $G$ distinct groups, with each group comprising $N=N_{\text{total}}/G$ frequency resources. Thus, a small-scale SCMA codebook with size of $N \times K$  can be employed within each group. Note that each SCMA codebook among different groups can be differentiated through constellation rotation \cite{SCMA1-4}. In this case, the system throughput can attain $GT$ based on \eqref{SCMA_4}, which ensures a great performance in the mMTC scenarios.  

The procedure of our proposed SCMA-empowered URA scheme towards mMTC scenarios is refined as follows.
Initially, the AP incorporates the group information into beacon signals and then broadcasts to all potential users. Subsequently, any user, who intends to initiate an uplink access, randomly selects a group index. Following the group index selection, the active user chooses the specific SCMA codebook associated with the selected group to convey its data over the uplink channel. Finally, at the AP, the proposed IC-first blind decoding strategy will be executed group by group.

The advantages of the proposed group RA strategy can be summarized as follows:
\begin{itemize}
\item{{\textit{Throughput improvement:}} By partitioning the frequency resources into multiple group, more SCMA codebooks becomes accessible to uplink users. Hence, such group random access strategy leads to the system throughput increase along with the number of groups, which is verified by simulation results.}
\end{itemize}

\begin{itemize}
\item{{\textit{Low implementation overhead:}} By assigning multiple small-scale SCMA codebook into different group, the above computational burden is distributed across various groups, which significantly reduces the overall complexity associated with codebook design and decoding. Hence, the implementation overhead is minimized, resulting in a cost-effective URA system.}
\end{itemize}

It is worth noting that under the limited available physical resources for random access combined with heavy real-time traffic loads, the system throughput still degrades due to frequent codebook collisions. In this case, a user barring mechanism is implemented to help maintain a satisfactory throughput performance, which will be discussed in the sequel.

\subsection{The application within satellite-terrestrial communications}
Given that the absence of complex coordination and scheduling mechanisms significantly reduce the signaling overhead and guarantee low access latency, ALOHA-based schemes are deemed as fundamental random access protocols in satellite-terrestrial communication systems \cite{gf-2}, \cite{Sat_1}-\cite{Sat_4}. Their simplicity and efficiency make ALOHA-based protocols particularly suitable for handling sporadic data traffic in satellite-based IoT scenarios.

As an enhanced version of ALOHA protocol, our proposed SCMA-empowered URA scheme inherits the aforementioned advantages while further improving the system performance. More importantly, the integration of SCMA technology boosts the system throughput, which meets the demand of massive access. In contrast to power-domain NOMA-based URA schemes \cite{Sat_3}, which rely on power disparities to distinguish users, the proposed SCMA-empowered URA scheme leverages unique SCMA codewords for user separation. This distinction is crucial in satellite-terrestrial communications, where long-distance signal transmission introduces unpredictable large-scale fading effects. Such fading may render received signal powers indistinguishable, which undermines the effectiveness of power-domain NOMA schemes. Conversely, SCMA can remain robust under these challenging channel conditions \cite{Sat_5}, making our proposed scheme as a more feasible solution for satellite communication systems.

\section{Simulation Results}
In this section, we present simulation results to show the performance of the the proposed SCMA-empowered URA scheme. Additionally, the accuracy of the derived theoretical results and the effectiveness of the proposed user barring mechanism are validated. In the following, two SCMA codebooks with different scales are adopted and their indicator matrices are represented by \eqref{Sim_1} and \eqref{Sim_2}, respectively. Besides, the user arrival rate follows the Poisson distribution, denoting as $\overline \lambda$ per time slot. For performance comparison, the multi-channel S-ALOHA system is utilized as the benchmark, which can be viewed as an example of the OMA-based URA scheme.
\begin{equation}\label{Sim_1}
\small
\setlength{\arraycolsep}{3pt}
{{\bf{F}}_{5 \times 10}} = \left[ {\begin{array}{*{20}{c}}
1&1&1&1&0&0&0&0&0&0\\
1&0&0&0&0&1&0&0&1&1\\
0&1&0&0&1&0&1&0&0&1\\
0&0&1&0&1&0&0&1&1&0\\
0&0&0&1&0&1&1&1&0&0
\end{array}} \right].
\end{equation}

\begin{equation}\label{Sim_2}
\small
\setlength{\arraycolsep}{3pt}
\begin{aligned}
{{\bf{F}}_{6 \times 15}} =\left[ {\begin{array}{*{20}{c}}
1&0&0&1&0&1&1&0&0&0&0&0&1&0&0\\
0&0&0&1&1&0&0&0&1&0&1&0&0&1&0\\
1&0&1&0&0&0&0&0&0&1&1&0&0&0&1\\
0&1&0&0&0&0&1&1&0&0&0&0&0&1&1\\
0&1&1&0&0&0&0&0&1&0&0&1&1&0&0\\
0&0&0&0&1&1&0&1&0&1&0&1&0&0&0
\end{array}} \right].
\end{aligned}
\end{equation}

\begin{figure}[t]
\centering{\includegraphics[width=0.38\textwidth,height=0.7\textwidth]{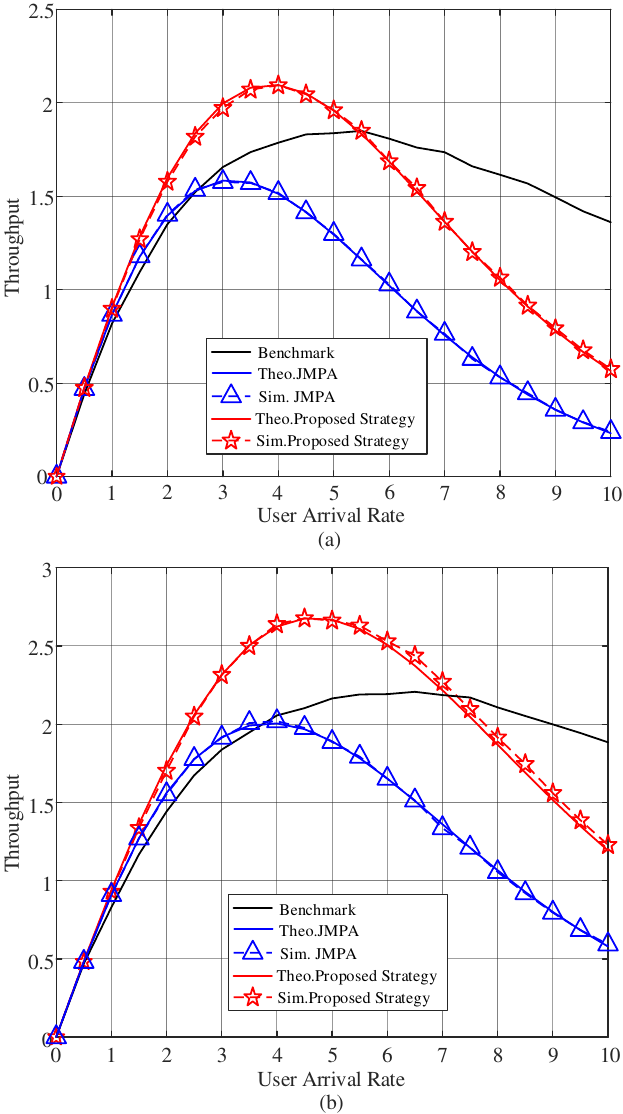}}
\caption{The throughput versus the varying user arrival rate. To be specific, SCMA codebooks with size of $5 \times 10$ and $6 \times 15$ are employed in the Fig. 3(a) and Fig. 3(b), respectively.}
\label{Throughput}
\vspace{-1em}
\end{figure}

Fig. \ref{Throughput} shows the throughput performance versus different user arrival rates. Noting that, the blue and red curves both correspond to the proposed SCMA-empowered URA scheme. The only difference is that the red curves adopt the proposed IC-first decoding strategy, while the blue ones use the JMPA method. As the benchmark, the performance of the multi-channel S-ALOHA system is represented by the black solid line.

Firstly, we focus on Fig. \ref{Throughput}(a), where the SCMA codebook with ${{\bf{F}}_{5 \times 10}}$ is adopted. In particular, from the indicator matrix \eqref{Sim_1}, we have $d_v = 2$, $d_f=4$ and $\theta= 200{\rm{\% }}$. From Fig. \ref{Throughput}(a), it can be seen that the proposed IC-first decoding strategy outperforms JMPA in the proposed SCMA-empowered S-ALOHA scheme. This is due to the fact that JMPA fails to work once codebook collisions take place. Conversely, the proposed IC-first approach is able to partially decode the superimposed codewords in the presence of codebook collisions, thus leading to a higher throughput. Moreover, the analytical results of using JMPA and the proposed decoding methods are plotted according to \eqref{SCMA_1} and \eqref{SCMA_4}, respectively. It is clear that the analytical expressions concur with the simulation results, which verifies the accuracy of the presented derivations. In terms of the throughput comparison between the proposed SCMA-empowered S-ALOHA scheme and the OMA-based scheme, one can notice that before reaching the throughput peak, the proposed scheme shows its superiority and more importantly, it enjoys a higher maximum throughput. To be specific, it can achieve a throughput of 2.1 when the user load $\overline \lambda = 4$. In the contrast, the maximum throughput of the OMA-based URA scheme equals to 1.839\footnotemark[1]. Such throughput improvement is attributed to the integration with SCMA technique, which allows the system to serve more users simultaneously than the number of available resources. However, when the system becomes congested and collisions accordingly become severe, the throughput degrades with the increasing user load. Especially, the throughput of the proposed scheme undergoes a sharper decline after reaching the peak value, compared with that of the OMA-based scheme. This is because each active user occupies multiple frequency resources with the use of SCMA codewords in the proposed scheme. Consequently, when collisions occur frequently, due to the overlapping nature of the non-orthogonal SCMA codewords, the performance degradation exacerbates more heavily compared to the OMA-based scheme, where each user can only occupy one orthogonal resource, resulting in a lower collision impact. This phenomenon underscores the importance of the user barring mechanism in the proposed SCMA-empowered S-ALOHA scheme, which plays a vital role in maintaining the throughput close to the peak value in overloading conditions.

\footnotetext[1]{The maximum throughput of the multichannel S-ALOHA scheme is linearly amplified with the number of subchannels \cite{NOSA-1}. For example, the maximum throughput of the multichannel S-ALOHA scheme with 5 subchannels equals to 5 times maximum throughput achieved by the conventional S-ALOHA scheme, i.e. $5e^{-1} \approx 1.839$.}

To further verify the performance and the accuracy of our analytical derivations for the proposed scheme, Fig. \ref{Throughput}(b) exhibits the throughput comparison when adopting the SCMA codebook with size $6 \times 15$ shown in \eqref{Sim_2}, whose overloading factor $\theta$ is $250{\rm{\% }}$. It can be seen that the red dash curve, plotted by the presented analytical expression in \eqref{SCMA_4}, can still closely align with the simulation result, despite of a slight deviation in high arrival rate region due to the approximation when deriving \eqref{SCMA_3}. Overall, the alignment between the analytical expression and simulation results demonstrate that the presented expression of the system throughput can be regarded as a reliable tool to conduct the real-time load estimation, thus facilitating the implementation of the user barring mechanism. Additionally, the maximum throughput with ${{\bf{F}}_{6 \times 15}}$ achieves 2.7, while that in the OMA-based URA is 2.2 (i.e., $6e^{-1}$) with the same available frequency resources. In other words, the throughput increases by $23{\rm{\% }}$ under this configuration, compared with $14{\rm{\% }}$ improvement with ${{\bf{F}}_{5 \times 10}}$. It reveals the fact that when the SCMA codebook with a higher overloading factor is adopted, a larger throughput enhancement can be achieved, since the system has the ability to accommodate more users concurrently utilizing the same resources.

\begin{figure}[t]
\centering{\includegraphics[width=0.38\textwidth]{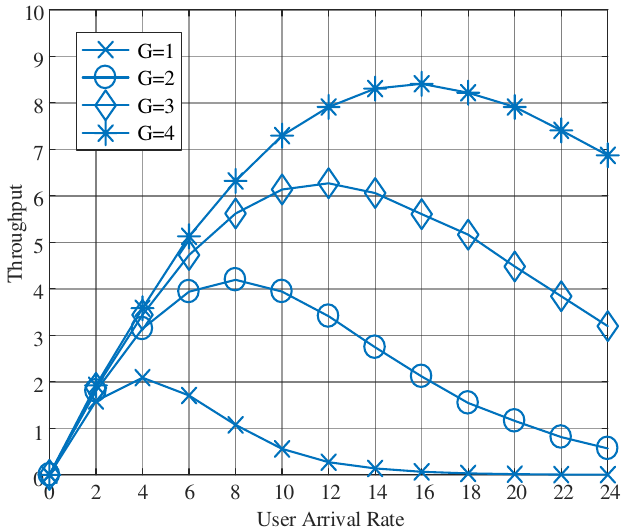}}
\caption{Throughput performance of the proposed SCMA-empowered URA scheme in massive access scenarios, where the SCMA codebook with size of $5 \times 10$ is used.}
\label{Th_group}
\vspace{-1em}
\end{figure}

In context of massive access scenarios, Fig. \ref{Th_group} provides the throughput performance of our proposed SCMA-empowered URA scheme with large user traffic load along with the increasing number of available SCMA codebooks. Each group employs an SCMA codebook of size $5 \times 10$.  As shown in Fig. \ref{Th_group}, the system throughput increases linearly with the number of available SCMA codebooks, demonstrating the effectiveness of the proposed group-based strategy in handling large-scale connectivity. In other words, as the number of available physical resources increases, our proposed SCMA-empowered scheme enables to accommodate more users simultaneously. Meanwhile, it is evident that when the real-time user load becomes excessively heavy, the throughput performance still suffers from a significant decline. This phenomenon can be attributed to the fact that, given a fixed number of available SCMA codebooks, there is an inherent limitation to the throughput performance they can support. In this case, to ensure that the system performance remains at a satisfactory level under such heavy-load conditions, the user barring scheme is implemented. The relevant simulation results are given in the following contents.


\begin{figure}[t]
\centering{\includegraphics[width=0.38\textwidth]{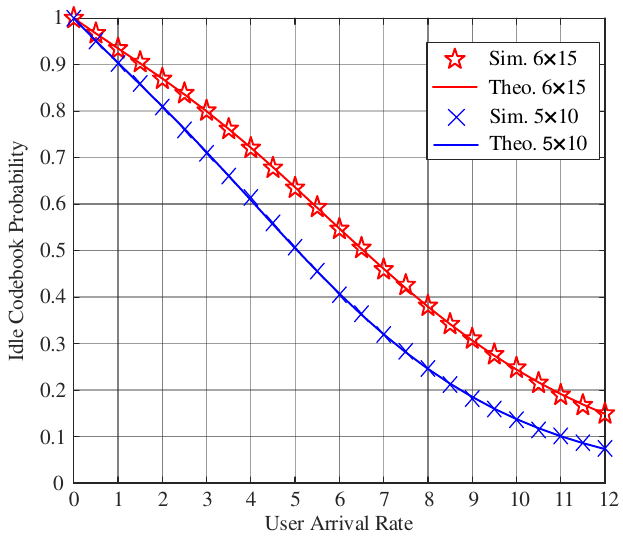}}
\caption{The idle codebook probability against different user arrival rates, where simulation and theoretical results are given with different size of SCMA codebooks.}
\label{idle}
\vspace{-1em}
\end{figure}

In Fig. \ref{idle}, the accuracy of the derived theoretical expression for ${P_\text{idle}}$ is demonstrated. Particularly, it can be observed that the presented theoretical result perfectly aligns with the simulation one under two different SCMA codebook configurations. In addition, with the rising traffic load, ${P_\text{idle}}$ gradually decreases from 1 to 0. It also indicates the monotonic decreasing nature of the ${P_\text{idle}}$ in relation to the user load. This property is also useful for determining the system's status, specifically whether it is in the underloading or overloading state, which enables an effective traffic load control. For ease of understanding, take the case with ${{\bf{F}}_{6 \times 15}}$ as an example. From Fig. \ref{idle}, the maximum throughput is achieved when $\overline \lambda = 4.5$. Accordingly, we can obtain ${P_\text{idle}^*} = {P_\text{idle}\left| {_{\lambda = 4.5}} \right.} = 0.68$. Thus, ${P_\text{idle}^*}=0.68$ is regarded as the load threshold to determine the system status. More precisely, when the real-time idle codebook probability exceeds 0.68, the system is identified as being in the underloading state. Conversely, when the real-time idle codebook probability falls below the threshold, the system is deemed to be in the overloading state.

\begin{figure}[t]
\centering{\includegraphics[width=0.4\textwidth,height=0.7\textwidth]{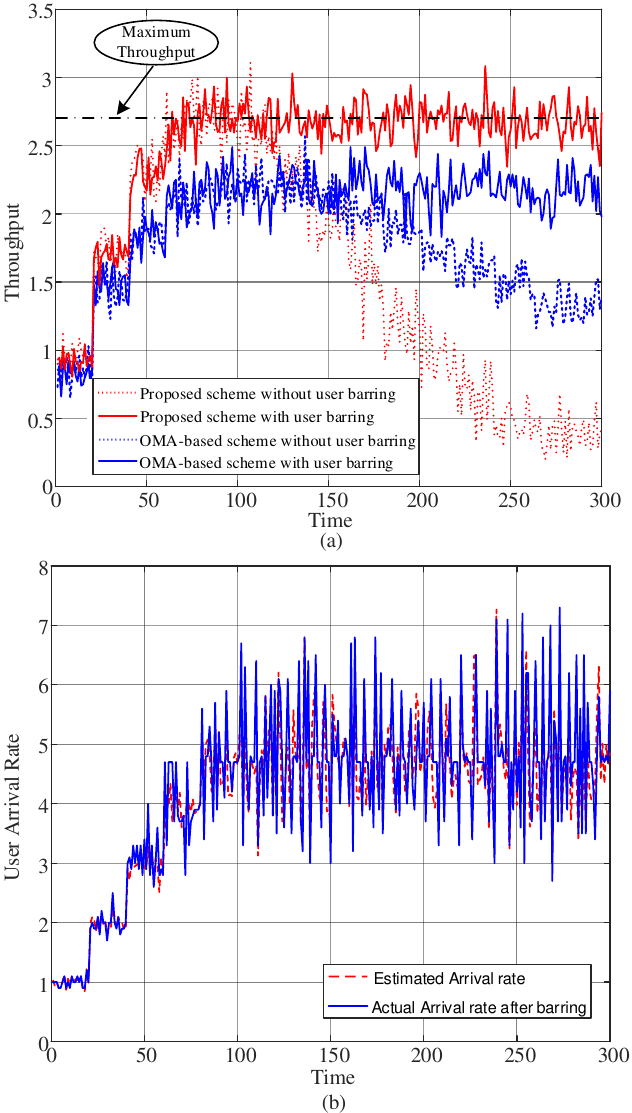}}
\caption{(a) shows the influences of the user barring mechanism on the throughput performance, while (b) provides the performance of the real-time traffic load estimation.}
\label{barring}
\vspace{-1.5em}
\end{figure}

Fig. \ref{barring} is provided to verify the effectiveness of the user barring mechanism, where the SCMA codebook with ${{\bf{F}}_{6 \times 15}}$ is employed. It is assumed that the transmission time lasts 300 frames and each frame includes 25 time slots. The user load estimation is executed after every whole frame transmission and the newly updated access probability $\alpha$ is broadcast to all users served by the AP. Note that the initial $\alpha$ is set as 1. Moreover, the user arrival rate is assumed to follow the Poisson distribution and remains fixed every 20 frames. For the first 20 frames, the arrival rate is $\overline \lambda = 1$. Subsequently, the arrival rate varies per 20 frames, which follows
\begin{equation}\label{Sim_3}
\overline \lambda = \frac{{\left\lceil {\text{Ti} - 20} \right\rceil }}{{20}} \times 1 + 1,\text{Ti} > 20,
\end{equation}
where $\text{Ti}$ denotes the index of the transmission frame. According to \eqref{Sim_3}, the intensity of $\overline \lambda$ adds 1 per 20 frames.

\footnotetext[2]{The uplink access protocol without the user barring mechanism implies that the access probability is fixed the value of 1. In other words, all users attempting uplink access are permitted to access the network in each transmission slot.}

Fig. \ref{barring}(a) shows the influences of the user barring mechanism on the throughput performance. Particularly, the performance comparisons with or without user barring mechanism\footnotemark[2] are given. It should be noted that the proposed SCMA-empowered URA scheme utilizes the IC-first decoding strategy. Shown in Fig. \ref{barring}(a), when the user arrival rate is light, the throughput of the proposed scheme increases with the rising load. However, when the traffic load becomes heavy, the throughput performance of the proposed scheme without user barring mechanism significantly degrades due to the frequent collisions. On the other hand, with the exploitation of the user barring mechanism, the throughput maintains close to the maximum value in the case of high traffic load. The reason is that the user barring mechanism manages the actual user load by adaptively adjusting the access probability $\alpha$, preventing the system from becoming excessively congested. Conversely, in the absence of the user barring mechanism, the system suffers from severe collisions, resulting in a degradation of the throughput. Moreover, compared with the OMA-based scheme, the proposed scheme with the aid of the user barring mechanism shows its superiority in terms of the throughput, regardless of fluctuations in user arrival rates. Specifically, recall that the prime advantage of the proposed SCMA-empowered S-ALOHA scheme is the higher maximum throughput, while the shortcoming is the sharp throughput performance degradation when the system becomes congested. From Fig. \ref{barring}(a), it can be concluded that the integration of the user barring mechanism effectively optimizes the performance of the SCMA-empowered URA scheme, ensuring that its major advantage, i.e., higher maximum throughput, can be fully realized while mitigating the potential drawbacks associated with overloading scenarios.

Fig. \ref{barring}(b) shows the results of the real-time traffic load estimation, where two main points can be observed. Firstly, the accuracy of the load estimation is verified since the blue and red curves match closely during the whole transmission periods. The robust performance of the load estimation method is essential, since it enables the user barring mechanism to make informed decisions on adjusting the access probability, leading to a satisfactory throughput optimization. Next, as depicted by the blue curve, the actual traffic load initially exhibits an increasing trend with time. However, as the system gradually becomes congested and user contention intensifies, the actual traffic load no longer shows an upward trend. This observation aligns with the findings from Fig. \ref{barring}(a). Specifically, in the overloading scenarios, the user barring mechanism adaptively adjusts the access probability and maintains the actual traffic load which is always close to the optimal value, preventing it from excessively increasing.

\section{Conclusion}
In this paper, we conceived a novel URA scheme, namely SCMA-empowered URA, aiming at supporting massive connectivity as well as ensuring low access latency. Firstly, we introduced an IC-first decoding strategy to address the superimposed signal at the AP, which is able to partially decode the received signal in the presence of codebook collisions. Then, the throughput of the proposed URA scheme has been analytically investigated when employing the IC-first decoding strategy and an accurate closed-form expression of the throughput has been derived. Additionally, the theoretical closed-form expression of the idle codebook probability was derived, which can be used to indicate the system congestion level. Subsequently, to resolve the potential drawback of rapid throughput degradation in congested situations, a user barring mechanism has been presented to reduce collisions and optimize the system throughput. Finally, simulation results verified the superiority of the proposed scheme in terms of the system throughput, compared with that of the conventional OMA-based URA scheme. The consistent alignment between the theoretical analysis and simulation results validates the accuracy of our presented theoretical framework, demonstrating its reliability in predicting the system performance. Additionally, the effectiveness of the user barring mechanism has also been successfully demonstrated. 

In conclusion, the proposed SCMA-empowered URA scheme constitutes a viable solution for future mMTC to achieve reduced signalling overhead and low access latency as well as enhanced throughput.

\begin{IEEEbiography}[{\includegraphics[width=1in,height=1.25in,clip,keepaspectratio]{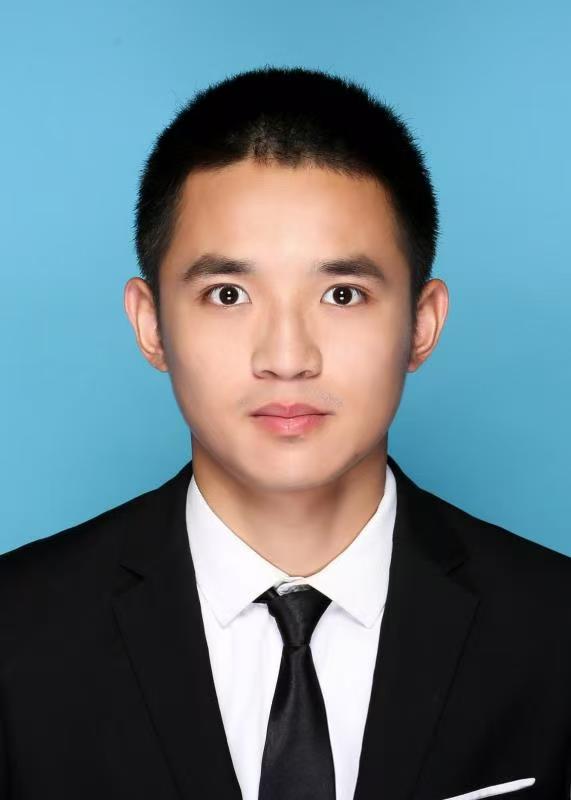}}]{Pengyu Gao}
 received the B.S. and M.S. degrees from UESTC, Chengdu, China, in 2016 and 2019, respectively, and the Ph.D. degree in 5G\&6G Innovation Centre of University of Surrey, U.K., in 2023. He is currently an associate researcher with Peng Cheng Laboratory, Shenzhen. His research interests include satellite communications, grant-free multiple access and new waveform design.
 \end{IEEEbiography}

 \begin{IEEEbiography}[{\includegraphics[width=1in,height=1.25in,clip,keepaspectratio]{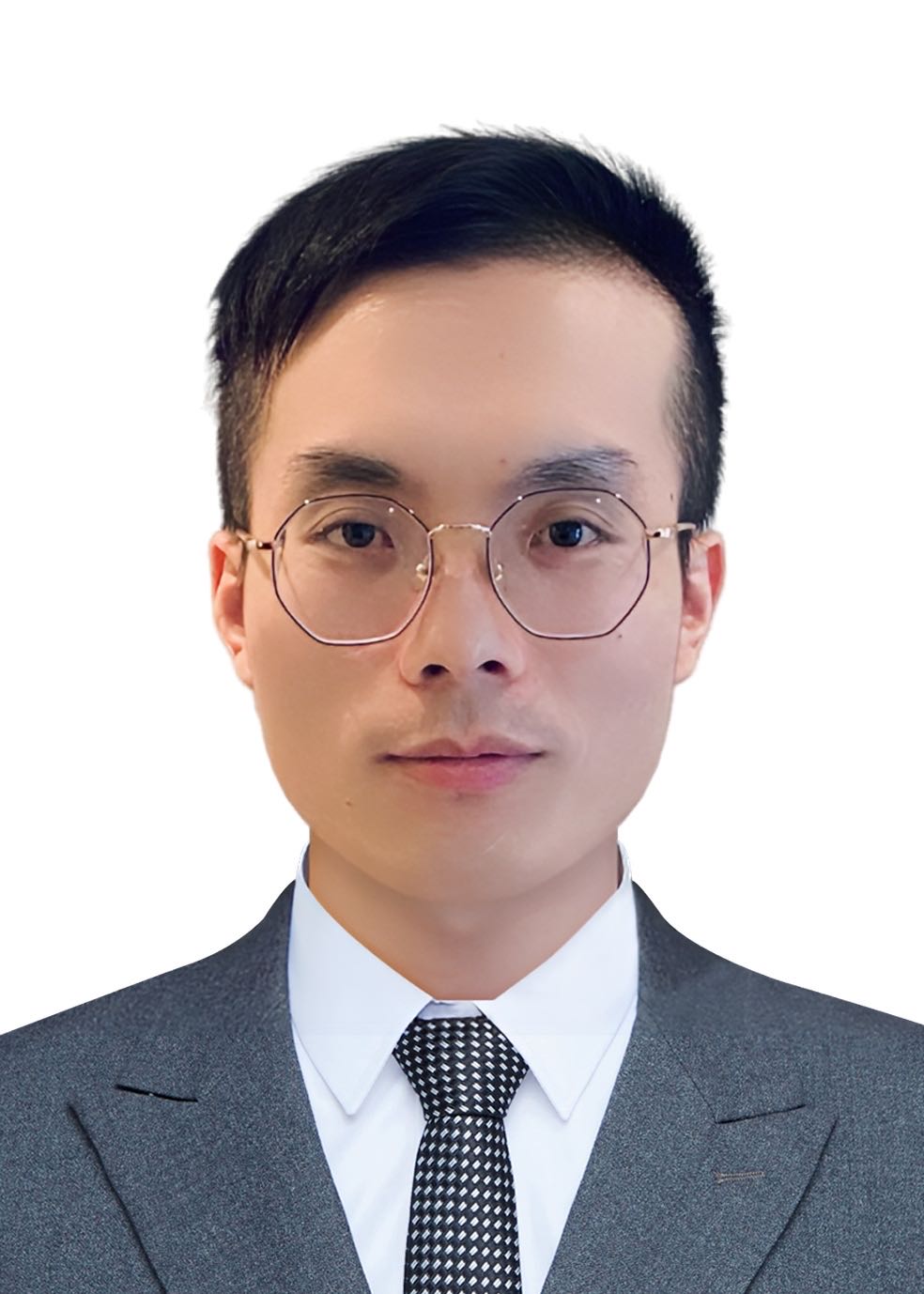}}]{Qu Luo}  is
 currently a Research Fellow in Wireless Communications in the 5GIC \& 6GIC, Institute for Communication Systems, University of Surrey, United Kingdom, where he received his Ph.D. degree  in 2023.  Prior to this, he worked  at the Huawei Technologies Company Ltd., Chengdu, China from 2019 to 2020.   He was also a  visiting  collaborative researcher at the  University of Essex in 2023.  He was a recipient of   the Exemplary Reviewer Awards of the IEEE Wireless Communication Letters from 2020 to 2023 and  the IEEE Communication Letters in 2022 and 2023, and   the Best Paper Awards of the IEEE CSPS in 2018 and the IWCMC in 2024. 
 He has also served as workshop organizer/co-organizer for  IEEE ICCC 25, IEEE SPAWC 25, and IEEE UCOM 25. 
 His  research interests include proof-of-concept physical layer design,   integrated  sensing and communication, non-orthogonal multiple access, random access, deep/machine learning in physical layer, and joint MAC layer and physical layer optimization. 
 \end{IEEEbiography}

 \begin{IEEEbiography}[{\includegraphics[width=1in,height=1.25in,clip,keepaspectratio]{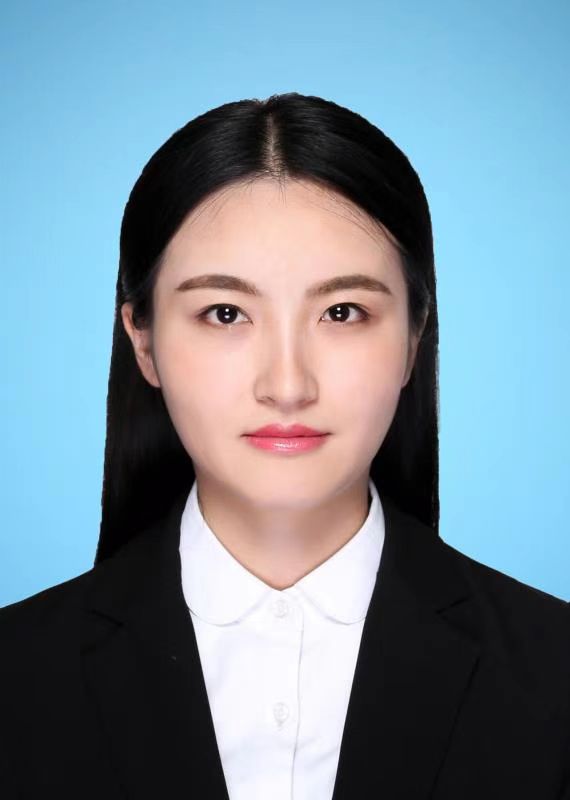}}]
{Jing Zhu} (Member, IEEE)
received the B.S and M.S. degree from the University of Electronic Science and Technology of China (UESTC), Chengdu, China, in 2016 and 2019, respectively, and the Ph.D. degree from the University of Surrey, U.K., in 2023. Her research interests include index modulation, flexible electronics, fluid antenna systems, RIS and massive MIMO. She was a recipient of the Exemplary Reviewer Awards of IEEE WIRELESS COMMUNICATION LETTERS in 2023 and IEEE COMMUNICATION LETTERS in 2021.
 \end{IEEEbiography}

\begin{IEEEbiography}[{\includegraphics[width=1in,height=1.25in,clip, keepaspectratio]{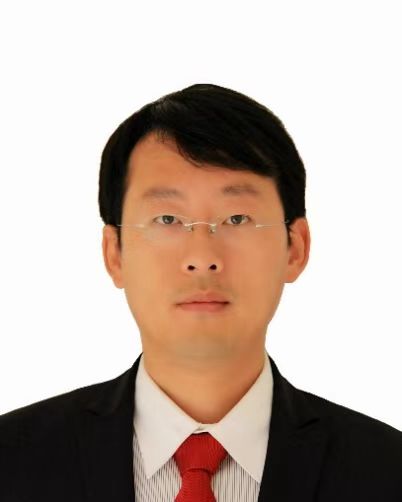}}]{Gaojie Chen}
(S'09 -- M'12 -- SM'18) received the B.Eng. and B.Ec. Degrees in electrical information engineering and international economics and trade from Northwest University, China, in 2006, and the M.Sc. (Hons.) and PhD degrees in electrical and electronic engineering from Loughborough University, Loughborough, U.K., in 2008 and 2012, respectively. After graduation, he took up academic and research positions at DT Mobile, Loughborough University, University of Surrey, University of Oxford and University of Leicester, U.K. He is a Professor and Associate Dean of the School of Flexible Electronics (SoFE), at Sun Yat-sen University, China, and a visiting professor at the 5GIC\&6GIC, University of Surrey, UK. His research interests include wireless communications, flexible electronics, satellite communications, the Internet of Things and secrecy communications. He received the Best Paper Awards from the IEEE IECON 2023, and the Exemplary Reviewer Awards of the {\scshape IEEE Wireless Communications Letters} in 2018, the {\scshape IEEE Transactions on Communications} in 2019 and the {\scshape IEEE Communications Letters} in 2020 and 2021; and Exemplary Editor Awards of the {\scshape IEEE Communications Letters}, {\scshape IEEE Wireless Communications Letters}, and {\scshape IEEE Transactions on Cognitive Communications Networking} in 2021, 2022, 2023 and 2024, respectively. He served as an Associate Editor for the {\scshape IEEE Journal on Selected Areas in Communications - Machine Learning in Communications} from 2021 to 2022. He serves as an Editor for the {\scshape IEEE Transactions on Wireless Communications}, {\scshape IEEE Transactions on Cognitive Communications Networking}, {\scshape IEEE Wireless Communications Letters}, and a Senior Editor for the {\scshape IEEE Communications Letters}, and a Panel Member of the Royal Society’s International Exchanges, UK.
\end{IEEEbiography}

  \begin{IEEEbiography}[{\includegraphics[width=1in,height=1.25in,clip,keepaspectratio]{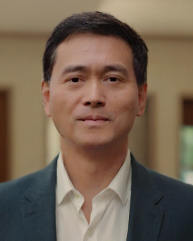}}]{Pei Xiao}  is a Professor in Wireless Communications in the Institute for Communication Systems (ICS) at University of Surrey. He is currently the technical manager of 5GIC/6GIC, leading the research team in the new physical layer work area, and coordinating/supervising research activities across all the work areas (https://www.surrey.ac.uk/institute-communication-systems/5g-6g-innovation-centre). Prior to this, he worked at Newcastle University and Queen’s University Belfast. He also held positions at Nokia Networks in Finland. He has published extensively in the fields of communication theory, RF and antenna design, signal processing for wireless communications, and is an inventor on over 15 recent patents addressing bottleneck problems in 5G/6G systems.
  \end{IEEEbiography}

  \begin{IEEEbiography}[{\includegraphics[width=1in,height=1.25in,clip,keepaspectratio]{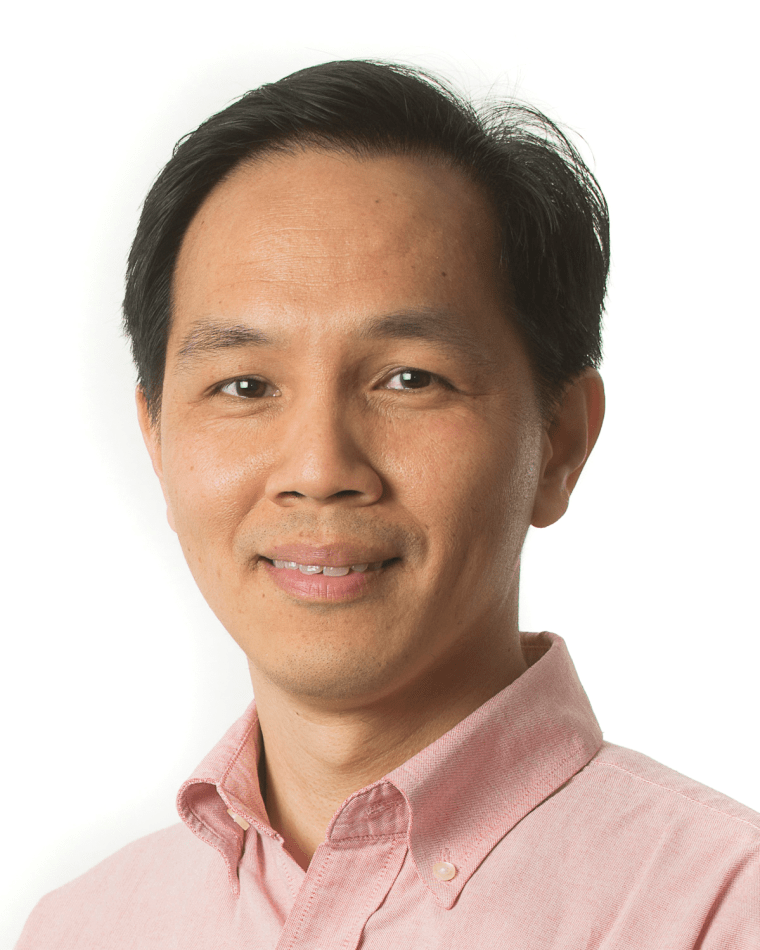}}]{Chuan Heng Foh}  (Senior Member, IEEE) received the M.Sc. degree from Monash University, Melbourne, VIC, Australia, in 1999, and the Ph.D. degree from the University of Melbourne, Melbourne, in 2002. After the Ph.D. degree, he spent six months as a Lecturer with Monash University. In December 2002, he joined Nanyang Technological University, Singapore, as an Assistant Professor, until 2012. He is currently an Associate Professor with the University of Surrey, Guildford, U.K. He has authored or co-authored more than 190 refereed papers in international journals and conferences. His research interests include protocol design, machine learning application, and performance analysis of various computer networks, including wireless local area networks, mobile ad-hoc and sensor networks, vehicular networks, the Internet of Things, 5G/6G networks, and open RAN. He served as the Vice Chair (Europe/Africa) for IEEE TCGCC, in 2015 and 2017. He is the Vice-Chair of the IEEE VTS Technical Committee on Mission Critical Communications. He is on the editorial boards of several international journals and a Senior Editor for IEEE Access.
  \end{IEEEbiography}
  
\end{document}